\documentclass[prl,showpacs,showkeys,twocolumn,amsmath,amssymb]{revtex4}
\usepackage{graphicx}
\usepackage{bm}
\usepackage[colorlinks,citecolor=blue,anchorcolor=red,menucolor=red,linkcolor=red,filecolor=red,runcolor=red,urlcolor=blue,frenchlinks=red]{hyperref}

\usepackage{ulem}
\usepackage{color}

\def\dab{\int^{\alpha_{max}}_{\alpha_{min}}d\alpha\int^{\beta_{max}}_{\beta_{min}}d\beta}
\def\qq{\langle\bar qq\rangle}

\def\GGa{\langle GG\rangle}
\def\GGb{\langle g_s^2GG\rangle}
\def\qGqa{\langle\bar qg_s\sigma\cdot Gq\rangle}
\def\qGqb{\langle\bar qGq\rangle}

\def\f(s){\left[(\alpha+\beta)m_c^2-\alpha\beta s\right]}
\def\non{\\ \nonumber}

\begin{document}
\title{Settling the $Z_c(4600)$ in the charged charmonium-like family}
%

\author{Hua-Xing Chen$^1$}
\author{Wei Chen$^2$}
\email{chenwei29@mail.sysu.edu.cn}
\affiliation{
$^1$School of Physics, Beihang University, Beijing 100191, China \\
$^2$School of Physics, Sun Yat-Sen University, Guangzhou 510275, China
}
\begin{abstract}
Very recently LHCb reported the evidence of a new charged charmonium-like structure in the $J/\psi \pi^-$ invariant mass spectrum near 4600 MeV. In this work we investigate this structure together with three other charged charmonium-like states, the $Z_c(3900)$, $Z_c(4020)$, and $Z_c(4430)$. Our results suggest that the two higher states can be established as the first radial excitations of the two lower ones, all of which have the quantum numbers $J^{PC}=1^{+-}$. We propose to search for more relationship among exotic hadrons in order to better understand them.
\end{abstract}
\pacs{12.39.Mk, 12.38.Lg, 12.40.Yx}
\keywords{tetraquark, diquark, QCD sum rules}
\maketitle
\pagenumbering{arabic}

$\\$
{\it Introduction.}
Since the discovery of the $X(3872)$ by the Belle Collaboration in 2003~\cite{Choi:2003ue}, there have been tens of charmonium-like $XYZ$ states observed in various particle experiments~\cite{pdg}, all of which are good multiquark candidates. Their relevant theoretical and experimental studies have significantly improved our understanding on the internal structure of (exotic) hadrons and the non-perturbative property of the strong interaction at the low energy region~\cite{Chen:2016qju,Klempt:2007cp,Lebed:2016hpi,Esposito:2016noz,Guo:2017jvc,Olsen:2017bmm}.
Some of these $XYZ$ states are not isolated, e. g., the $Y(4260)$ can decay radiatively to the $X(3872)$~\cite{Ablikim:2013dyn}. These connections can give important hints on their properties.
It is thus important to search for possible connections among different $XYZ$ states, which shall shed light on our understanding of their underlying properties.

To date, the charged charmonium-like $Z_c$ family already has at least five members: the $Z_c(3900)$~\cite{Ablikim:2013mio,Liu:2013dau}, $Z_c(4020)/Z_c(4025)$~\cite{Ablikim:2013wzq,Ablikim:2013emm}, $Z_c(4100)$~\cite{Aaij:2018bla}, $Z_c(4200)$~\cite{Chilikin:2014bkk}, and $Z_c(4430)$~\cite{Choi:2007wga,Aaij:2014jqa}, etc.
Very recently, the evidence of another charged charmonium-like structure was reported by the LHCb Collaboration in the $J/\psi \pi^-$ invariant mass spectrum near 4600 MeV, after performing an angular analysis of the $B^0 \rightarrow J/\psi K^+ \pi^-$ decay~\cite{Aaij:2019ipm}. We temporarily denote it as $Z_c(4600)$. Actually, the Belle Collaboration has also studied the $\bar B^0 \rightarrow J/\psi K^- \pi^+$ process in 2014~\cite{Chilikin:2014bkk}, where they found the evidence for the $Z_c(4430)$, that is the effect of destructive interference in the $J/\psi \pi^+$ mass spectrum near 4485 MeV, instead of a peaking structure near 4600 MeV. Hence, it is crucial to verify whether there exists the $Z_c(4600)$ or not experimentally.

%
\begin{figure*}[hbt]
\begin{center}
\includegraphics[width=1\textwidth]{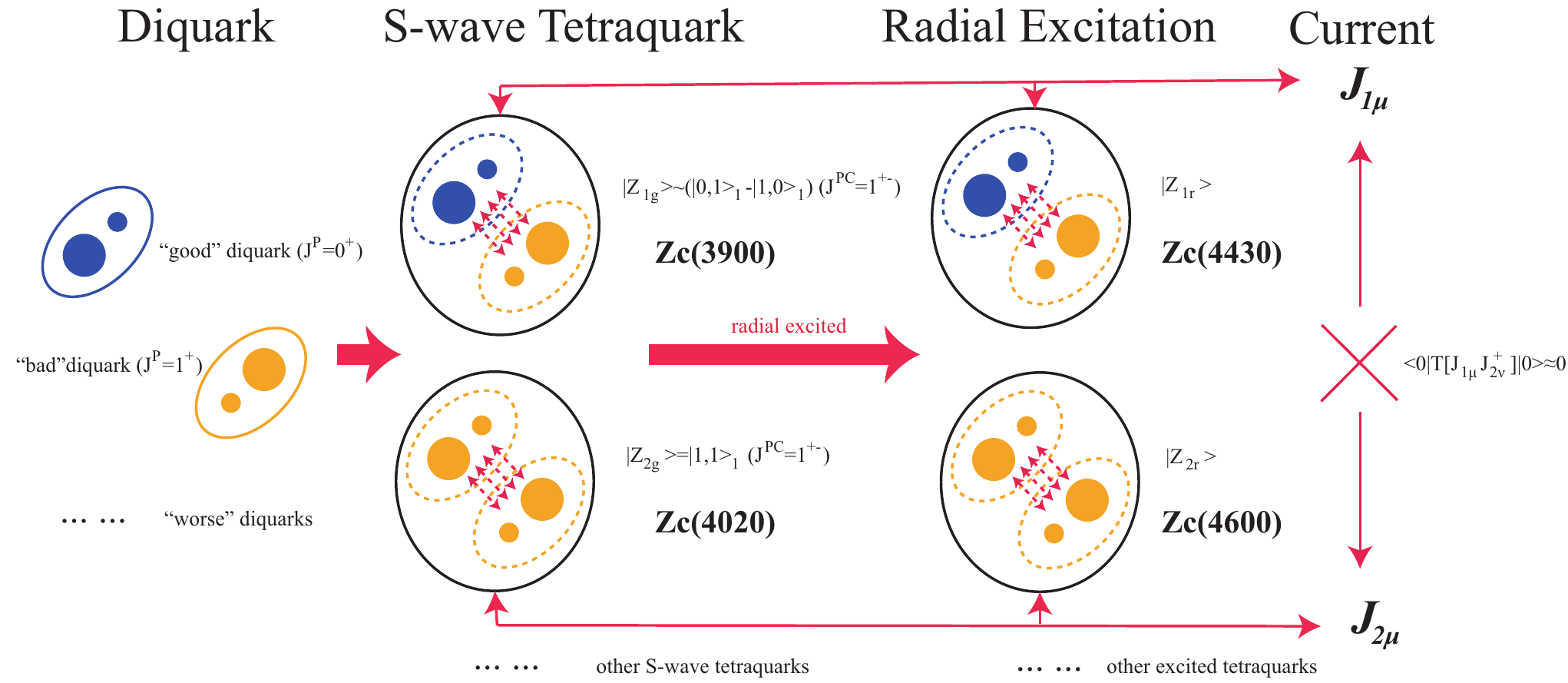}
\caption{Possible explanations of the $Z_c(3900)$, $Z_c(4020)$, $Z_c(4430)$, and $Z_c(4600)$ as a whole, supported by a) the phenomenological analyses within the diquark model~\cite{Maiani:2014aja}, and b) the QCD sum rule analyses performed in the present study.}
\label{fig:diagram}
\end{center}
\end{figure*}
%

Recalling that one possible explanation of the $Z_c(4430)$ is to interpret it as the first radial excitation of the $Z_c(3900)$~\cite{Maiani:2007wz,Maiani:2014aja,Ebert:2008kb,Nielsen:2014mva,Wang:2014vha,Agaev:2017tzv}  (see reviews~\cite{Chen:2016qju,Klempt:2007cp,Lebed:2016hpi,Esposito:2016noz,Guo:2017jvc,Olsen:2017bmm} for more possible explanations), which seems to be reasonable because their mass difference is about 591 MeV, very close to the mass difference between the $\psi(2S)$ and $\psi(1S)$. Accordingly, it is natural to consider the $Z_c(4600)$ as the first radial excitation of some other $Z_c$ state, such as the $Z_c(4020)$. Moreover, in the diquark model~\cite{Maiani:2014aja} there exist two $S$-wave tetraquark states with $J^{PC} = 1^{+-}$, which can be used to explain the $Z_c(3900)$ and $Z_c(4020)$. Therefore, it seems to exist a close relationship among the $Z_c(3900)$, $Z_c(4020)$, $Z_c(4430)$, and $Z_c(4600)$.

In this paper we study the $Z_c(4600)$ together with the $Z_c(3900)$, $Z_c(4020)$, and $Z_c(4430)$. We first use the diquark model proposed in Ref.~\cite{Maiani:2014aja} to perform a phenomenological analysis, and then apply the method of QCD sum rules~\cite{Shifman:1978bx,Reinders:1984sr} to study their relationship. The results obtained from these two approaches both suggest the following explanations to be possible, as illustrated in Fig.~\ref{fig:diagram}: a) the $Z_c(4430)$ and $Z_c(4600)$ are the first radial excitations of the $Z_c(3900)$ and $Z_c(4020)$, respectively; b) all of them are composed of $S$-wave charmed diquarks and antidiquarks; c) all of them have the quantum numbers $J^{PC} = 1^{+-}$ (for neutral charge states). Since we have studied them as a whole, several assumptions/predictions are used/made at the same time, and we propose the experimentalists to: a) verify whether the $Z_c(4600)$ exists or not, b) determine its quantum numbers as well as those of the $Z_c(4020)$, c) search for their partner states with different quark contents. Especially, both theoretical and experimental studies on the relationship of exotic hadrons are intriguing research topics.

$\\$
{\it Phenomenological analyses within the diquark model.} Firstly, let us phenomenologically estimate the masses of the $Z_c(3900)$, $Z_c(4020)$, $Z_c(4430)$, and $Z_c(4600)$, assuming that they are composed of one charmed diquark ($cq$) and one charmed antidiquark ($\bar c \bar q$). To do this we use the ``type-II'' diquark-antidiquark model proposed in Ref.~\cite{Maiani:2014aja}, and we refer to Refs.~\cite{Maiani:2014aja,Maiani:2004vq,Ali:2017wsf,Maiani:2017kyi} for its detailed discussions. In this model, the $S$-wave tetraquarks can be written in the spin basis as $|s, \bar s\rangle_J$, where $s = s_{qc}$ and $\bar s = s_{\bar q \bar c}$ are the charmed diquark and antidiquark spins, respectively.

There are two $S$-wave diquarks: the ``good'' diquark with $J^P = 0^+$ and the ``bad'' diquark with $J^P = 1^+$ (other diquarks are ``worse'')~\cite{Jaffe:2004ph}, so $s/\bar s$ can be either 0 or 1.
Consequently, there are altogether seven $S$-wave tetraquark states, denotes as $|X;J^{PC}\rangle$:
\begin{eqnarray}
\nonumber
|X_0 ; 0^{++} \rangle &=& | 0, 0 \rangle_{0} \, ,
\non |X_0^\prime ; 0^{++} \rangle &=& | 1, 1 \rangle_0 \, ,
\non \sqrt2 \times |X_1 ; 1^{++} \rangle &=& | 0, 1 \rangle_1 + |1, 0 \rangle_1 \, ,
\\ \sqrt2 \times |Z_1 ; 1^{+-} \rangle &=& | 0, 1 \rangle_1 - |1, 0 \rangle_1 \, ,
\non |Z_2 ; 1^{+-} \rangle \equiv |Z_1^\prime ; 1^{+-} \rangle &=& | 1, 1 \rangle_1 \, ,
\non |X_2 ; 2^{++} \rangle &=& | 1, 1 \rangle_2 \, .
\end{eqnarray}
Especially, there are two tetraquark states with $J^{PC} = 1^{+-}$: $|Z_1\rangle$ and $|Z_2\rangle$. Note that the notation $|Z_1^\prime \rangle$ is used in Ref.~\cite{Maiani:2014aja}, while $|Z_2\rangle$ is used in this paper for convenience.
In the ``type-II'' diquark-antidiquark model, their masses can be evaluated through~\cite{Maiani:2014aja}
\begin{eqnarray}
M_X = 2 M_{cq} + 2\kappa_{cq} ( {\bm s}_c \cdot {\bm s}_q + {\bm {\bar s}}_c \cdot {\bm {\bar s}}_q ) \, ,
\end{eqnarray}
where $M_{cq}$ is the effective charmed diquark mass, and ${\bm s}_c/{\bm s}_q/{\bm {\bar s}}_c/{\bm {\bar s}}_q$ are the quark and antiquark spins. According to this mass formula:
\begin{enumerate}

\item Identifying $Z_1 \rightarrow Z_c(3900)$ and $Z_2 \rightarrow Z_c(4020)$, we can use the experimental masses of the $Z_c(3900)$ and $Z_c(4020)$ to obtain
\begin{equation}
M_{cq} \approx  1978 ~ {\rm MeV} \, .
\end{equation}

\item With the above diquark mass, we can use the Cornell potential for charmonia~\cite{Eichten:1978tg,Eichten:1979ms,Ikhdair:2003ry,Chen:2015dig}
\begin{equation}
V(r) = -{0.47 \over r} + r \times 0.19~{\rm GeV}^2 \, ,
\end{equation}
to roughly estimate the radial excitation energy between charmed diquark and antiquark to be about 581 MeV.

\end{enumerate}
Accordingly, the masses of the first radial excitations of the $Z_c(3900)$ and $Z_c(4020)$ are about 4467 MeV and 4605 MeV, respectively. These two values are well consistent with the experimental masses of the $Z_c(4430)$ and $Z_c(4600)$, suggesting that the latter two can be interpreted as the first radial excitations of the former two. Again, we refer to Fig.~\ref{fig:diagram} for an illustration of this picture.

$\\$
{\it Constructions of tetraquark interpolating currents.} In the following we shall use the method of QCD sum rules~\cite{Shifman:1978bx,Reinders:1984sr} to investigate the above interpretations.
Similar to the above non-relativistic case, there are two $S$-wave diquark fields:
\begin{eqnarray}
&& \epsilon^{abc} q_a^T C \gamma_5 c_b \, ,~~~J^P = 0^+\, ,
\\ && \epsilon^{abc} q_a^T C \gamma_\mu c_b \, ,~~~J^P = 1^+\, ,
\end{eqnarray}
where $a/b$ are color indices. We can combine them to construct the tetraquark current corresponding to $|Z_1 ; 1^{+-} \rangle ={1\over\sqrt2} \left( | 0, 1 \rangle_1 - |1, 0 \rangle_1 \right)$:
\begin{eqnarray}
\nonumber
J_{1\mu} &=& \left( \epsilon^{abc} q_a^T C \gamma_5 c_b \right) \times \left( \epsilon^{a^\prime b^\prime c} \bar q_{a^\prime} \gamma_\mu C \bar c_{b^\prime}^T \right)
\\ \nonumber && ~~~~~ -  \left( \epsilon^{abc} q_a^T C \gamma_\mu c_b \right) \times \left( \epsilon^{a^\prime b^\prime c} \bar q_{a^\prime} \gamma_5 C \bar c_{b^\prime}^T \right)
\\ \nonumber &=& q_a^T C \gamma_5 c_b \left( \bar q_{a} \gamma_\mu C \bar c_{b}^T - \bar q_{b} \gamma_\mu C \bar c_{a}^T \right)
\\ && ~~~~~ -  q_a^T C \gamma_\mu c_b \left( \bar q_{a} \gamma_5 C \bar c_{b}^T - \bar q_{b} \gamma_5 C \bar c_{a}^T \right) \, .
\end{eqnarray}
We need to use the tensor diquark field $\epsilon^{abc} q_a^T C \sigma_{\mu\nu} \gamma_5 c_b$ to construct another tetraquark current with $J^{PC} = 1^{+-}$
\begin{eqnarray}
\nonumber
J_{2\mu} &=& \left( \epsilon^{abc} q_a^T C \gamma^\nu c_b \right) \times \left( \epsilon^{a^\prime b^\prime c} \bar q_{a^\prime} \sigma_{\mu\nu} \gamma_5 C \bar c_{b^\prime}^T \right)
\\ \nonumber && ~~~~~ -  \left( \epsilon^{abc} q_a^T C \sigma_{\mu\nu} \gamma_5 c_b \right) \times \left( \epsilon^{a^\prime b^\prime c} \bar q_{a^\prime} \gamma^\nu C \bar c_{b^\prime}^T \right)
\\ \nonumber &=& q_a^T C \gamma^\nu c_b \left( \bar q_{a} \sigma_{\mu\nu} \gamma_5 C \bar c_{b}^T - \bar q_{b} \sigma_{\mu\nu} \gamma_5 C \bar c_{a}^T \right)
\\ && ~~~~~ -  q_a^T C \sigma_{\mu\nu} \gamma_5 c_b \left( \bar q_{a} \gamma^\nu C \bar c_{b}^T - \bar q_{b} \gamma^\nu C \bar c_{a}^T \right)\, .
\end{eqnarray}
In principle, the tensor diquark field $\epsilon^{abc} q_a^T C \sigma_{\mu\nu} \gamma_5 c_b$ can couple to both $J^P = 1^+$ and $1^-$ channels. However, its positive-parity component $\epsilon^{abc} q_a^T C \sigma_{ij} \gamma_5 c_b$ ($i, j=1, 2, 3$) gives the dominant contribution to $J_{2\mu}$. Hence, the tetraquark interpolating current $J_{2\mu}$ with $J^{PC} = 1^{+-}$ corresponds to $|Z_2 ; 1^{+-} \rangle = | 1, 1 \rangle_1$.

The tetraquark currents $J_{1\mu}$ and $J_{2\mu}$ have been used in Ref.~\cite{Chen:2010ze} to perform QCD sum rule analyses, and the masses extracted are about $4.02 \pm 0.09$ MeV and $4.14 \pm 0.09$ MeV, respectively. These two values are slightly larger than the experimental masses of the $Z_c(3900)$ and $Z_c(4020)$. Such results may imply that $J_{1\mu}$ and $J_{2\mu}$ can couple to both the ground-state tetraquarks as well as their radial excitations. In the present study we shall consider both the contributions of ground states and their radial excitations.

$\\$
{\it QCD sum rule analyses at the quark-gluon level.} Since the relation between physical states and their relevant interpolating currents is complicated, for example, it is possible that $J_{1\mu}$ couples to both the $Z_c(3900)$ and $Z_c(4020)$ as well as their radial excitations, we need to study $J_{1\mu}$ and $J_{2\mu}$ themselves as well as their mixing in order to achieve a more reliable analysis, given that we do not know how to evaluate the mixing between the $Z_c(3900)$ and $Z_c(4020)$ theoretically. We refer to Refs.~\cite{Chen:2018kuu,Cui:2019roq} for detailed discussions.

We investigate both the diagonal and off-diagonal correlation functions ($i/j=1/2$):
%
\begin{eqnarray}
\Pi_{ij;\mu\nu}(q^2) &\equiv& i \int d^4x e^{iqx} \langle 0 | T J_{i\mu}(x) { J_{j\nu}^\dagger } (0) | 0 \rangle
\label{def:pi}
\\ \nonumber &=& ( {q_\mu q_\nu \over q^2} - g_{\mu\nu} ) \Pi_{ij}(q^2) + {q_\mu q_\nu \over q^2} \Pi^{(0)}_{ij}(q^2) \, .
\label{def:pi1}
\end{eqnarray}
%
In QCD sum rule studies we express $\Pi_{ij}(q^2)$ in the form of the dispersion relation with spectral functions $\rho_{ij}(s)$:
%
\begin{equation}
\Pi_{ij}(q^2)=\int^\infty_{4 m_c^2}\frac{\rho_{ij}(s)}{s-q^2-i\varepsilon}ds \, ,
\label{eq:disper}
\end{equation}
%
which can be further transformed into
%
\begin{equation}
\Pi_{ij}(M_B^2) = \int^{\infty}_{4 m_c^2}~e^{-s/M_B^2}~\rho_{ij}(s)~ds \, ,
\label{eq_borel}
\end{equation}
%
by using the Borel transformation. Here $M_B$ is the Borel mass.

\begin{figure}[hbt]
\begin{center}
\includegraphics[width=0.4\textwidth]{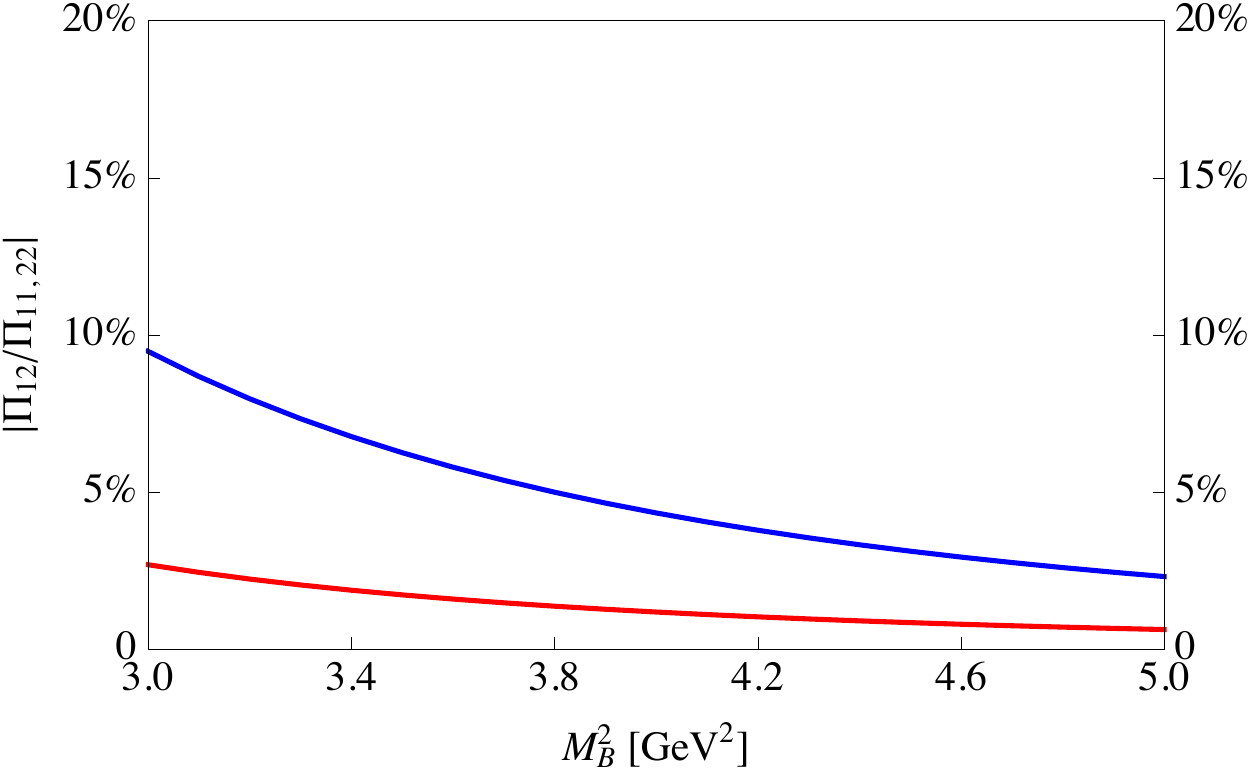}
\end{center}
\caption{$\left|{\Pi_{12}(M_B^2)\over\Pi_{11}(M_B^2)}\right|$ (blue) and $\left|{\Pi_{12}(M_B^2)\over\Pi_{22}(M_B^2)}\right|$ (red), as functions of the Borel mass $M_B^2$.
}
\label{fig:offdiagonal}
\end{figure}

At the quark and gluon level, we can calculate $\rho_{ij}(s)$ using the method of operator product expansion (OPE). In the present study we have done this up to dimension eight condensates. The diagonal spectral densities $\rho_{11}(s)$ and $\rho_{22}(s)$ have been calculated and given in Ref.~\cite{Chen:2010ze}, and the off-diagonal spectral density $\rho_{12}(s)$ is
\begin{eqnarray*}
\nonumber
\rho_{12}^{\GGa}(s)&=&\frac{\GGb}{36864\pi^6}\dab\non&&\Bigg\{\frac{(1-\alpha-\beta)^2\left[m_c^2(4\alpha-14\beta-7)-21\alpha\beta
s\right]}{\alpha^2\beta^2}\non&&
-\frac{6\left[m_c^2(4\alpha+4\beta-5)-7\alpha\beta
s\right]}{\alpha\beta}\Bigg\}\non &&\f(s)\, , \non
\rho_{12}^{\qGqb}(s)&=&\frac{m_c\qGqa}{96\pi^4}\dab
\non &&
\frac{\left[3m_c^2(\alpha+\beta)-5\alpha\beta
s\right]}{\beta}\, , \\
\rho_{12}^{\qq\qGqb}(s)&=&\frac{\qq\qGqa}{96\pi^2}\int_0^1\alpha\theta(s-\tilde{m}_c^2) d\alpha\, ,
\end{eqnarray*}
where $\theta(s-\tilde{m}_c^2)$ is a step function and
\begin{align*}
\nonumber
\alpha_{max}&=\frac{1+\sqrt{1-4m_c^2/s}}{2}\, , \, \,
\alpha_{min}=\frac{1-\sqrt{1-4m_c^2/s}}{2}\, , \, \,
\\
\beta_{max}&=1-\alpha\, ,  \beta_{min}=\frac{\alpha
m_c^2}{\alpha s-m_c^2}\, , \, \, \tilde{m}_c^2=\frac{m_c^2}{\alpha(1-\alpha)}\, .
\end{align*}
To perform numerical analyses, we use the values listed in Ref.~\cite{Chen:2010ze} for the charm quark mass and various condensates (see also Refs.~\cite{pdg,Yang:1993bp,Narison:2002pw,Gimenez:2005nt,Jamin:2002ev,Ioffe:2002be,Ovchinnikov:1988gk,Ellis:1996xc}).
We show $\Pi_{12}(M_B^2)$ in Fig.~\ref{fig:offdiagonal} as a function of the Borel mass $M_B^2$, compared with $\Pi_{11}(M_B^2)$ and $\Pi_{22}(M_B^2)$. It shows that $J_{1\mu}$ and $J_{2\mu}$ only weakly correlate to each other, and thus can not strongly couple to the same physical state.

$\\$
{\it QCD sum rule analyses at the hadron level.}
In the present study we assume that $J_{1\mu}$ couples to both the ground-state tetraquark $Z_{1g}$ and its first radial excitation $Z_{1r}$:
\begin{equation}
\langle 0| J_{1\mu} | Z_{1g} \rangle = f_{1g}\epsilon_{\mu} \, , \, \langle 0| J_{1\mu} | Z_{1r} \rangle = f_{1r}\epsilon_{\mu} \, ,
\label{eq:coupling}
\end{equation}
while $J_{2\mu}$ couples to other states. After inserting Eqs.~(\ref{eq:coupling}) into the two-point correlation function (\ref{def:pi}), we obtain its expression at the hadron level:
\begin{equation}
\Pi_{11}(q^2) = {f_{1g}^2 \over M_{1g}^2 - p^2} + {f_{1r}^2 \over M_{1r}^2 - p^2} + \cdots \, ,
\label{pi:hadron}
\end{equation}
where $\cdots$ denote the contribution from other higher states (the continuum); $M_{1g}$ and $M_{1r}$ are the masses of $Z_{1g}$ and $Z_{1r}$, respectively. Again, we perform the Borel transformation to Eq.~(\ref{pi:hadron}) and obtain
\begin{equation}
\Pi_{11}(M_B^2) = f_{1g}^2 e^{- M_{1g}^2 / M_B^2 } + f_{1r}^2 e^{ - M_{1r}^2 / M_B^2 } + \cdots \, .
\label{correlation:hadron}
\end{equation}

One important assumption in the QCD sum rule approach is the quark-hadron duality, which ensures the equivalence of the correlation functions obtained at the quark-gluon level and the hadron level. Accordingly, we assume the contribution from the continuum states can be approximated well by the OPE spectral density above a threshold value $s_0$, and obtain
\begin{eqnarray}
\Pi_{11}(M_B^2, s_0) &=& f_{1g}^2 e^{- M_{1g}^2 / M_B^2 } + f_{1r}^2 e^{ - M_{1r}^2 / M_B^2 }
\label{eq:sumrule}
\\ \nonumber &=& \mathcal{L}_0(M_B^2, s_0) \, ,
\end{eqnarray}
where we have used the notation
\begin{equation}
\mathcal{L}_n(M_B^2, s_0) \equiv \int^{s_0}_{4 m_c^2}~e^{-s/M_B^2}~\rho_{11}(s)~s^n~ds \, .
\end{equation}
To extract $M_{1g}$ and $M_{1r}$, one can differentiate Eq.~(\ref{eq:sumrule}) with respect to $(-{1 / M_B^2})$ up to three times~\cite{Wang:2014vha}:
\begin{eqnarray}
\nonumber M_{1g}^2 f_{1g}^2 e^{- M_{1g}^2 / M_B^2 } + M_{1r}^2 f_{1r}^2 e^{ - M_{1r}^2 / M_B^2 } &=& \mathcal{L}_1(M_B^2, s_0) \, ,
\\ \nonumber M_{1g}^4 f_{1g}^2 e^{- M_{1g}^2 / M_B^2 } + M_{1r}^4 f_{1r}^2 e^{ - M_{1r}^2 / M_B^2 } &=& \mathcal{L}_2(M_B^2, s_0) \, ,
\\ \nonumber M_{1g}^6 f_{1g}^2 e^{- M_{1g}^2 / M_B^2 } + M_{1r}^6 f_{1r}^2 e^{ - M_{1r}^2 / M_B^2 } &=& \mathcal{L}_3(M_B^2, s_0) \, .
\end{eqnarray}
The unknown parameters $M_{1g}$, $f_{1g}$, $M_{1r}$, and $f_{1r}$ can be obtained by solving the above three equations together with Eq.~(\ref{eq:sumrule}). Focusing on the hadron masses, both $M_{1g}$ and $M_{1r}$ can
satisfy the following equation
\begin{equation}
M^4 - b M^2 + c = 0 \, ,
\end{equation}
where
\begin{equation}
b = {\mathcal{L}_3 \mathcal{L}_0 - \mathcal{L}_2 \mathcal{L}_1 \over \mathcal{L}_2\mathcal{L}_0 - \mathcal{L}_1\mathcal{L}_1} \, , \,
c = {\mathcal{L}_3 \mathcal{L}_1 - \mathcal{L}_2 \mathcal{L}_2 \over \mathcal{L}_2\mathcal{L}_0 - \mathcal{L}_1\mathcal{L}_1} \, .
\end{equation}
After carefully investigating a) the OPE convergence, b) the pole contribution, and c) the mass dependence on the two free parameters $M_B$ and $s_0$, we obtain reliable QCD sum rule results in the regions $3.5$~GeV$^2 < M_B^2< 4.5$~GeV$^2$ and $21$~GeV$^2 < s_0< 23$~GeV$^2$, where the hadron masses are extracted to be
\begin{eqnarray}
M_{1g} &=& 3.85^{+0.22}_{-0.17} ~{\rm GeV} \, ,
\\ M_{1r} &=& 4.53^{+0.16}_{-0.10} ~{\rm GeV} \, .
\end{eqnarray}
Here the central values correspond to $M_B^2 = 4.0$~GeV$^2$ and $s_0 = 22$~GeV$^2$, and the uncertainties are due to the Borel mass $M_B$, the threshold value $s_0$, and various QCD condensates.
These two mass values are consistent with the experimental masses of the $Z_c(3900)$ and $Z_c(4430)$.

Similarly, we assume that $J_{2\mu}$ couples to another ground-state tetraquark $Z_{2g}$ as well as its first radial excitation $Z_{2r}$:
\begin{equation}
\langle 0| J_{2\mu} | Z_{2g} \rangle = f_{2g}\epsilon_{\mu} \, , \, \langle 0| J_{2\mu} | Z_{2r} \rangle = f_{2r}\epsilon_{\mu} \, ,
\label{eq:coupling2}
\end{equation}
whose masses are extracted to be
\begin{eqnarray}
M_{2g} &=& 4.05^{+0.23}_{-0.20} ~{\rm GeV} \, ,
\\ M_{2r} &=& 4.70^{+0.29}_{-0.13} ~{\rm GeV} \, .
\end{eqnarray}
Here the central values correspond to $M_B^2 = 4.5$~GeV$^2$ and $s_0 = 23$~GeV$^2$. These two mass values are in good agreement with the experimental masses of the $Z_c(4020)$ and $Z_c(4600)$.
Our investigations can support the picture illustrated in Fig.~\ref{fig:diagram} that the $Z_c(4430)$ and $Z_c(4600)$ can be well interpreted
as the first radial excitations of the $Z_c(3900)$ and $Z_c(4020)$, respectively, consistent with the phenomenological analyses within the diquark model~\cite{Maiani:2014aja}.

$\\$
{\it Summary and Discussions.} Very recently, the LHCb Collaboration reported the evidence of a new charged charmonium-like structure in the $J/\psi \pi^-$ invariant mass spectrum near 4600 MeV~\cite{Aaij:2019ipm}. In this
work we investigate it together with three other charged charmonium-like structures, the $Z_c(3900)$, $Z_c(4020)$, and $Z_c(4430)$. We first estimate their masses within the diquark model proposed in Ref.~\cite{Maiani:2014aja}, and then study them using the method of QCD sum rules. Especially, in sum rule analyses we have used two weakly correlated interpolating currents, $J_{1\mu}$ and $J_{2\mu}$, which independently couple to different tetraquark states. The results from both two approaches suggest the following possible explanations, as illustrated in Fig.~\ref{fig:diagram}:
\begin{itemize}

\item The $Z_c(3900)$ and $Z_c(4020)$ are $S$-wave tetraquark states with $J^{PC} = 1^{+-}$. The $Z_c(3900)$ contains one ``good'' diquark with $J^P = 0^+$ and one ``bad'' diquark with $J^P = 1^+$, while the $Z_c(4020)$ contains two ``bad'' diquarks with $J^P = 1^+$.

\item The $Z_c(4430)$ and $Z_c(4600)$ are the first radial excitations of the $Z_c(3900)$ and $Z_c(4020)$, respectively. They also have the quantum numbers $J^{PC} = 1^{+-}$.

\end{itemize}
Note that there are many other possible explanations for the $Z_c(3900)$, $Z_c(4020)$, and $Z_c(4430)$, and we refer interested readers to Refs.~\cite{Chen:2016qju,Klempt:2007cp,Lebed:2016hpi,Esposito:2016noz,Guo:2017jvc,Olsen:2017bmm} for more discussions.

In the present study we have studied the $Z_c(3900)$, $Z_c(4020)$, $Z_c(4430)$, and $Z_c(4600)$ as a whole, and used/made several assumptions/predictions at the same time. To verify these assumptions/predictions, we propose the experimentalists to: a) further study the structure observed in the $J/\psi \pi^-$ mass spectrum near 4600 MeV to verify whether there exists a genuine charged charmonium-like state, b) determine its quantum numbers as well as those of the $Z_c(4020)$, and c) search for their partner states with the quark contents $[cs][\bar c \bar s]$ and $[bq][\bar b \bar q]$, etc. Especially, we propose to establish more connections among exotic hadrons in order to better understand them. Both theoretical and experimental studies on this topic are intriguing, which can further improve our understanding on the internal structure of (exotic) hadrons and the non-perturbative property of the strong interaction at the low energy region.

%
\section*{Acknowledgments}
%

We thank Prof. Luciano Maiani and Prof. Shi-Lin Zhu for useful discussions.
This project is supported by
the National Natural Science Foundation of China under Grants No. 11722540,
the Fundamental Research Funds for the Central Universities,
and the Chinese National Youth Thousand Talents Program.

%

%

\end{document}